\documentclass[aps,prd,superscriptaddress,tightenlines,nofootinbib,eqsecnum, 10pt,notitlepage,a4paper,preprintnumbers]{revtex4-2}

\usepackage{amsmath}
\usepackage{amsfonts}
\usepackage{amssymb}
\usepackage{bm}
\usepackage[colorlinks]{hyperref}
\usepackage{mathrsfs}
\usepackage{graphicx}
\usepackage{empheq}
\usepackage{ulem}
\usepackage{tensor}
\usepackage[usenames]{color}
\definecolor{darkgreen}{rgb}{0,0.5,0}
\hypersetup{urlcolor=darkgreen}
\usepackage[capitalize]{cleveref}

\allowdisplaybreaks

\DeclareSymbolFontAlphabet{\mathrsfs}{rsfs}
\DeclareMathAlphabet{\mathcal}{OMS}{cmsy}{m}{n}

\newcommand{\dd}{\mathrm{d}}

\newcommand{\Mpl}{M_\mathrm{Pl}}
\newcommand{\nn}{\nonumber}
\newcommand{\hp}{\hat{\partial}}
\newcommand{\bg}{\bar{g}}
\newcommand{\bGa}{\bar{\Gamma}}
\newcommand{\bn}{\bar{\nabla}}
\newcommand{\bR}{\bar{R}}

\newcommand{\gh}{\mathfrak{h}}

\begin{document}

\title{Gauss-Bonnet dynamical compactification scenarios\\
and their ghosts in the tensor sectors}

\author{Antonio De Felice}\email{antonio.defelice@yukawa.kyoto-u.ac.jp}
\affiliation{Center for Gravitational Physics and Quantum Information,\\
Yukawa Institute for Theoretical Physics, Kyoto University, 606-8502, Kyoto, Japan}

\author{Fran\c{c}ois \textsc{Larrouturou}}\email{francois.larrouturou@obspm.fr}
\affiliation{SYRTE, Observatoire de Paris, Universit\'{e} PSL, CNRS, Sorbonne Universit\'{e}, LNE,
61 avenue de l’Observatoire, 75014 Paris, France}

\preprint{YITP-24-163}

\begin{abstract}

In a cosmological context, the Einstein-Gauss-Bonnet theory contains, in $d+4$ dimensions, a dynamical compactification scenario in which the additional dimensions settle down to a configuration with a constant radion/scale factor.
Sadly however this work demonstrates that such a quite appealing framework is plagued by instabilities, either from the background configuration's unsteadiness or the ghostly behaviors of the tensorial perturbations.
New and stable solutions are found by relaxing one of the hypotheses defining the original compactification scenario.
However, such configurations do not respect the current bounds on the speed of propagation of gravitational waves, and thus have to be discarded.
Those results thus advocate for a comprehensive study of compactification scenarios in the Gauss-Bonnet framework, their stability, and the effects of matter inclusion.
\end{abstract}

\maketitle

\section{Introduction}

General Relativity (GR) is a conceptually extremely simple theory (if not the simplest) describing the dynamics of spacetime only by its Ricci scalar and a constant term.
It can be seen as the first term of a tower of theories, constructed with increasingly high powers of the curvature tensors.
Requiring the equations of motion for the only fundamental variables in the theory, the metric tensor, to be coordinate-diffeomorphism invariant, local, and, at most, second-order, drastically constrains the form of those higher-order terms, as investigated by D.~Lovelock~\cite{Lovelock:1971yv}. 
Notably, the second-order term is simply given by the Gauss-Bonnet combination of the Riemann tensor.
In four dimensions, those terms do not contribute to the dynamics, and Lovelock's theories reduce to GR.\\

In spacetimes with more than four dimensions, higher-order curvature terms contribute significantly to the dynamics, leading to a fascinating phenomenon in cosmological contexts: spontaneous compactification. Specifically, when the four-dimensional physical sub-manifold follows the standard de Sitter dynamics, the additional spatial dimensions "wrap up" into a compactified configuration.
Such scenarios are attractive because they naturally modify the canonical cosmological framework without introducing new fundamental fields beyond the metric.\\

This effect has been extensively studied in the context of the Einstein-Gauss-Bonnet framework~\cite{Chirkov:2014nua, Chirkov:2017fkx, Pavluchenko:2017qne, Chirkov:2018qlz, Chirkov:2019qbe, Chirkov:2020bnd} and higher-order Lovelock theories~\cite{Pavluchenko:2012rc, Chirkov:2015kja, Chirkov:2018xrd}. Furthermore, the behavior of scalar fields within such compactified configurations has also been investigated~\cite{Pozdeeva:2019agu}. Notably, when scalar fields are non-minimally coupled to curvature, the compactification mechanism can induce spontaneous symmetry breaking~\cite{Chirkov:2024rbv}.\\

As such a compactification scenario is really interesting \emph{per se}, one should ensure its viability before undertaking deep cosmological studies.
The linear stability against departures from the background configuration, via purely background homogeneous perturbations, has been studied, \emph{e.g.}~in~\cite{Chirkov:2020bnd}, and ranges where the solution is an attractor in the parameter space have been determined.
However, to our knowledge, no further stability studies have been pursued.
The present work tackles this open question, notably by studying the tensorial modes, and demonstrates that none of the compactified solutions are stable.\\

This unstable behavior is a strong motivation to search for other compactification scenarios, by relaxing at least one of the hypotheses on which the original mechanism is based.
Doing a minimal change to the initial scenario, an appealing and stable solution is found.
However, this solution is incompatible with the observational bounds on the speed of gravitational waves on cosmological scales and must be discarded.
It would thus be interesting to search the whole parameter space to exhaust the possible compactification scenarios and investigate their stability.
Such exciting work is left for future studies.\\

This work is organized as follows.
Sec.~\ref{sec_backg} presents the model and the compactification scenario for any number of dimensions, but all those configurations are proven unstable in Sec.~\ref{sec_instab}.
As such derivations can be somewhat arid, Sec.~\ref{sec_example} concretely illustrate them by focusing on the case of a $(4+4)$ dimensional configuration with a different approach, hence providing also an independent check of the previous computations.
Then, Sec.~\ref{sec_change} presents a tentative to evade those instabilities, by a minimal departure from the hypothesis of the original compactification scenario.
A conclusion and discussion is to be found in Sec.~\ref{sec_concl}.
App.~\ref{app_lengthy} collects expressions that are too long to be displayed in the text, and App.~\ref{app_pert_tens} details the computations leading to the quadratic perturbation of the theory.

\section{Dynamical compactification in cosmological frameworks}\label{sec_backg}

The present work aims to investigate the stability of the cosmological compactification scenarios in the Einstein-Gauss-Bonnet framework.

\subsection{The framework}

Following~\cite{Chirkov:2020bnd}, we thus consider a multidimensional Einstein-Gauss-Bonnet theory with the cosmological constant in the vacuum. 
Given a $(4+d)$-dimensional\footnote{As the cases $d=0$ or $1$ are trivial, we do not consider them here.} manifold $\mathcal{M}$ equipped with a metric tensor $g_{\mu\nu}$, the dynamics is fixed by the action
\begin{equation}\label{eq_action_base}
    \mathcal{S} = \frac{\Mpl^2}{2\,\ell_0^d}\int\!\!\dd^{d+4}x\sqrt{-g}\bigg\lbrace R-2\Lambda + \alpha\, \mathcal{G}\bigg\rbrace\,,
\end{equation}
where $\Mpl$ is the usual Planck mass, $\ell_0$ a constant introduced for dimensional reasons, $\alpha$ and $\Lambda$ are constants, $R$ is the Ricci scalar and $\mathcal{G}$ is the Gauss-Bonnet term, defined as~\cite{Lanczos:1938sf,Lovelock:1971yv,Deruelle:2003ck}
\begin{equation}\label{eq_GB_definition}
    \mathcal{G} = R_{\mu\nu\rho\lambda}R^{\mu\nu\rho\lambda}-4\,R_{\mu\nu}R^{\mu\nu}+ R^2\,,
\end{equation}
with $R_{\mu\nu\rho\sigma}$ and $R_{\mu\nu}$ respectively the Riemann and Ricci tensors computed from the metric $g_{\mu\nu}$. In the following, greek letters denote indices associated with the manifold $\mathcal{M}$ and we use a mostly positive signature for the metric.
The constant $\ell_0$ has been introduced so that $\Mpl$ is the usual, four-dimensional, Planck mass. As it will play no role in the following, we set arbitrarily, $\ell_0 = 1$.

\subsection{Cosmological background equations and compactified solutions}

To study compactified solutions in a cosmological context, let us consider that the manifold $\mathcal{M}$ is a product of two sub-manifolds, $\mathcal{M}_4 \times \mathcal{M}_d$, and use the cosmologically motivated diagonal-by-block line element
\begin{equation}\label{eq_lineelement_cosmobckg}
    \dd s^2  =-N^2(t)\,\dd t^2 + a^2(t)\,\hat{\delta}_{ij}\dd x^i \dd x^j + b^2(t) \,\omega_{AB}\,\dd x^A \,\dd x^B\,,
\end{equation}
where $\{x^i\}$ spans the usual three-dimensional space, $\{x^A\}$ spans the $d$-dimensional sub-manifold $\mathcal{M}_d$, $N(t)$ is a lapse function, $a(t)$ and $b(t)$ are two scale factors, $\hat{\delta}_{ij}$ is a flat metric and $\omega_{AB}$ is a $d$-dimensional constant-curvature metric, \emph{i.e.} its Riemann tensor reads
\begin{equation}\label{eq_RABCD_bckg}
    R[\omega]_{ABCD} = \kappa\, \big( \omega_{AC}\omega_{BD} - \omega_{AD}\omega_{BC}\big)\,.
\end{equation}
The curvature constant, $\kappa$, can be positive, negative, or vanishing.
From the scale factors, one can define associated Hubble-Lema\^{i}tre factors
\begin{equation}\label{eq_Hubble_Lemaitre_factors}
    H \equiv \frac{\dot{a}}{Na}
    \qquad \text{and} \qquad
    L \equiv \frac{\dot{b}}{Nb}\,.
\end{equation}
Injecting the metric~\eqref{eq_lineelement_cosmobckg} in the action~\eqref{eq_action_base} and varying it \emph{wrt.}\ $\{N, a,b\}$ gives equations of motion that are quite long and will not be displayed here but are presented in the App.~\ref{app_eom_bckg}.\\

Then, as we seek compactified solutions, we take the long-time limit of a de Sitter configuration for the four-dimensional spacetime and a constant radius $d$-dimensional sub-manifold. This compactification scenario amounts to taking
\begin{equation}\label{eq_compactification_ansatz}
    N(t) \to 1\,,\qquad
    H(t) \to H_0 \qquad \text{and}\qquad b(t) \to b_0\,.
\end{equation}
Naturally, the constant $b_0$ has to be strictly positive and, as we require a de Sitter configuration of the physical sub-manifold, we impose $H_0 > 0$.
As expected, the equations of motion are redundant, and the constraint~\eqref{eq_eom_bckg_N} turns out to be proportional to the equation of motion for $a(t)$~\eqref{eq_eom_bckg_a}, once the parametrization~\eqref{eq_compactification_ansatz} is injected.
A compactification scenario hence exists if the obtained system of two equations admits a solution for the variables $H_0$ and $b_0$ such that both are strictly positive.

\subsubsection{Flat extra-dimensional sub-manifolds require fine tuning}\label{sec:flatEDfinetuning}

Considering a flat extra-dimensional sub-manifold, \emph{i.e.} taking $\kappa = 0$, the system of equations trivially reduces to
\begin{equation}
    \Lambda = 3 H_0^2\,\qquad\text{and} \qquad \alpha = - \frac{1}{4H_0^2}\,.
\end{equation}
Therefore, the two parameters of the action are linked by $\alpha\Lambda = - 3/4$, which reveals a strong fine-tuning.
Hence, we will not consider a flat extra-dimensional manifold in the following.

\subsubsection{Constant-curvature extra-dimensional sub-manifolds allow dynamical compactification}

In the case of a curved extra-dimensional sub-manifold, $\kappa$ can be normalized to be $\pm 1$.
Denoting
\begin{equation}\label{eq_adim_var}
    \beta \equiv \alpha\,H_0^2\,,
    \qquad
    \lambda \equiv \frac{\Lambda}{H_0^2}
    \qquad \text{and}\qquad
    B_0 \equiv b_0 H_0\,,
\end{equation}
the system~\eqref{eq_eom_bckg} reduces to
\begin{subequations}
\begin{align}
    &
   \lambda = 3 + \frac{d(d-1)\,\kappa}{2B_0^2} + \frac{d(d-1)(d-2)(d-3)}{2 B_0^4}\,\beta + \frac{6d(d-1)\,\kappa}{B_0^2}\,\beta\,,\\
    &
   \lambda = 6(1+2\beta) + \frac{(d-1)(d-2)\,\kappa}{2B_0^2} + \frac{(d-1)(d-2)(d-3)(d-4)}{2 B_0^4}\beta + \frac{12(d-1)(d-2)\,\kappa}{B_0^2}\,\beta\,,
\end{align}
\end{subequations}
or equivalently
\begin{subequations}
\label{eq_lambda_beta_sol}
\begin{align}
    &
   \lambda = \frac{d(d-1)\,\kappa}{4\,B_0^2}  +\frac{72\,B_0^4+6(d-1)(d-24)\kappa\,B_0^2+6(d-1)(7d-12)}{4\big[6\,B_0^4+3(d-1)(d-4)\,\kappa\,B_0^2-(d-1)(d-2)(d-3)\big]}\,\\
    &
   \beta = \frac{\big[(d-1)\kappa-3B_0^2\big]\,B_0^2}{2\big[6\,B_0^4+3(d-1)(d-4)\,\kappa\,B_0^2-(d-1)(d-2)(d-3)\big]}\,.
\end{align}
\end{subequations}
It is easy to see that $d=2,3,4$ are specific cases and that $\lambda$ and $\beta$ are well defined (we recall that, by definition, $B_0 \neq 0$), excepted possibly at $B_0 = B_0^\pm$, where
\begin{equation}\label{eq_B0pm}
    B_0^\pm = \sqrt{- \frac{(d-1)(d-4)}{4}\,\kappa \pm \frac{1}{12}\,\sqrt{3d(d-1)(3d^2-19d+32)}}\,.
\end{equation}
When $\kappa=1$, there exists, therefore, one and only one value of $B_0$ for which $\beta$ is not defined. For $\kappa=-1$, $\beta$ is defined everywhere for $d=2,3$, and there exists one, and only one value of $B_0$ for which it is not defined when $d\geq 4$.\\

When $\kappa = \pm 1$, the Gauss-Bonnet term thus allows a dynamical compactification scenario, except for the possibility for a single configuration of the parameters, determined by Eq.~\eqref{eq_B0pm}.

\section{Instability of the compactified solutions}\label{sec_instab}

The compactification mechanism presented in the previous section has to be stable against gravitational perturbations to be viable.
This section is devoted to demonstrating that it is not.
To do so, we start by investigating the regions of parameter space in which the background solution is an attractor (which amounts to studying its stability against scalar breathing perturbations of the metric).
Then, we show that in such regions of parameter space, at least one of the tensor modes is a ghost.
Even if we work without matter fields, such ghosts render the theory non-viable, as it couples to healthy gravitational modes through non-linearities. 

\subsection{Attractor behaviour of the background}\label{sec_attractor}

The analysis of the attractor behavior of the compactified solutions has been performed in~\cite{Chirkov:2020bnd}.
For coherence with the study of tensor perturbation performed hereafter, we repeat such analysis, taking a slightly different approach from the one cited above.
Our conclusions are naturally in full agreement with those of~\cite{Chirkov:2020bnd}.\\

Setting without loss of generality the lapse function to unity, the dynamical system~\eqref{eq_eom_bckg} is of first-order in $H$ and second-order in $b$. 
Due to its redundancy, one can, however, transform it into a system of three first-order equations, by introducing the auxiliary variable $u \equiv \dot{b}$.
Perturbing the system in linear order by introducing
\begin{equation}
    H = H_0 \big( 1 + \delta H\big)\,,
    \qquad
    b = b_0 + \frac{\delta b}{H_0}
    \qquad\text{and}\qquad
    u = \delta u\,,
\end{equation}
using the dimensionless variables~\eqref{eq_adim_var}, it can be recast as
\begin{equation}\label{eq_pert_bckg_eqlin}
    H_0\,\begin{pmatrix}
        \delta \dot{H} \\
        \delta \dot{b} \\
        \delta \dot{u} \\
    \end{pmatrix}
    = M_0^\kappa\,
    \begin{pmatrix}
        \delta H \\
        \delta b \\
        \delta u \\
    \end{pmatrix}
    + \mathcal{O}(2)\,,
\end{equation}
where the matrix $M_0^\kappa$, evaluated on-shell using the compactification scenario~\eqref{eq_compactification_ansatz} and the background solution~\eqref{eq_lambda_beta_sol}, is presented in Eq.~\eqref{eq_M0kappa}.
It depends only on $\{B_0,\kappa,d\}$, and the background solution is an attractor if its three eigenvalues have negative real parts.\\

It is to be noted that the structure of the eigenvalues is given by
\begin{equation}
    \mu_0^\kappa = - 3
    \qquad\text{and}\qquad
    \mu_\pm ^\kappa = - \frac{3}{2} \pm \sqrt{\frac{\Pi^\kappa}{\Sigma^\kappa}}\,,
\end{equation}
where $\{\Pi^\kappa, \Sigma^\kappa\}$ are polynomials of $B_0$ of order 10, and read
\begin{subequations}
\begin{align}
    \Pi^\kappa = \
    & \nn
    324\,B_0^{10} 
    - 18(25d-14)\,\kappa\,B_0^8 
    - 12(8d^2-101d+111)B_0^6 \\
    & \nn
    + 18(16d-63)(d-1)(d-2)\,\kappa\,B_0^4
    -6(9d-32)(2d-3)(d-1)(d-2)B_0^2\\
    & \label{eq_def_Pikappa}
    + 8(2d-3)(d-1)^2(d-2)(d-3)\,\kappa\,,\\
    \Sigma^\kappa = \
    & \bigg.
    12\,B_0^2\big[ 2B_0^2 - (d-2)\kappa\big]\big[6B_0^6- 12(d-1)\,\kappa\,B_0^4+6(d-1)^2B_0^2 -d(d-1)(d-2)\kappa\big]\,.
\end{align}
\end{subequations}
The real parts of $\mu_0^\kappa$ and $\mu_-^\kappa$ are always negative, hence we need to investigate the behavior of the last eigenvalue, $\mu_+^\kappa$.

\subsubsection{When $d=2$, the background solution is never an attractor}\label{sec_background_stab_d2}

As will be clear in the following, the case $d=2$ is quite specific, so it is treated apart from the others.\\

Indeed, it simply comes
\begin{equation}
    \mu_+^\kappa = \frac{3}{2}\Bigg[\sqrt{1 + \frac{32}{27(1-\kappa\,B_0^2)^2}}-1\Bigg]\,.
\end{equation}
Note that when $\kappa= 1$, the eigenvalue is not defined at $B_0 = 1$, which is nothing but the value of $B_0^+$, see Eq~\eqref{eq_B0pm}.
It is easy to see that $\mu_+^\kappa$ is always strictly positive, hence the background solution is never an attractor.

\subsubsection{When $d \geq 3$ and $\kappa=-1$, the background solution is always an attractor}\label{sec_background_stab_kappaminus}

For extra-dimensional manifolds with negative curvatures, the polynomial $\Sigma^{(-)}$ has no real roots excepted 0, hence the eigenvalue $\mu_+^{(-)}$ is well defined.
Then, a study of the zeros of the derivative of the ratio $\Pi^{(-)}/\Sigma^{(-)}$ shows that it is a monotonic function for $B_0>0$. 
Expanding towards infinity
\begin{equation}
    \frac{\Pi^{(-)}}{\Sigma^{(-)}} = \frac{9}{4} - \frac{5(d-2)}{2\,B_0^2} + \mathcal{O}\left( B_0^{-4}\right)\,,
\end{equation}
reveals that, for $d \geq 3$, it is an increasing function on $\mathbb{R}^+_\star$, bounded by its value at infinity, $9/4$. 
It is thus clear that the eigenvalue $\mu^{(-)}_+$ always bears a negative real part, and so that when $\kappa = -1$, the compactified solution is an attractor.

\subsubsection{When  $d \geq 3$ and $\kappa=1$, there exist ranges of the parameter space where the background solution is an attractor}\label{sec_background_stab_kappaplus}

For extra-dimensional manifolds with positive curvatures, the polynomial $\Sigma^{(+)}$ has two strictly positive roots when $d \geq 5$, ordered as $B_0^{(\Sigma,1)} < B_0^{(\Sigma,2)}$ and presented in App.~\ref{app_rootB}.
Then, the derivative of the ratio $\Pi^{(+)}/\Sigma^{(+)}$ also has two strictly positive zeros, but those are smaller than $B_0^{(\Sigma,1)}$. Hence the ratio is monotonic in a range that starts before $B_0^{(\Sigma,1)}$ and extends up to infinity.
Expanding the ratio near infinity,
\begin{equation}
    \frac{\Pi^{(+)}}{\Sigma^{(+)}} = \frac{9}{4} + \frac{5(d-2)}{2\,B_0^2} + \mathcal{O}\left( B_0^{-4}\right)\,,
\end{equation}
reveals that this monotonic behavior is a decreasing one.
Therefore, when approaching $B_0^{(\Sigma,1)}$ and $B_0^{(\Sigma,2)}$ by below, the ratio goes to the negative infinity and so $\sqrt{\Pi^{(+)}/\Sigma^{(+)}}$ becomes at first smaller than $3/2$ before turning into a purely imaginary number.
Solving for $\Pi^{(+)}/\Sigma^{(+)}=9/4$ gives the lower bound of the range on which the real parts of the eigenvalue $\mu_+^{(+)}$ is strictly negative, namely $B_0^{(\Pi,1)}$ and $B_0^{(\Pi,2)}$, that are presented in App.~\ref{app_rootB}.
Hence, for $d \geq 5$ and $\kappa = 1$, there exist two ranges on which the background solution is an attractor.\\

When $d=3$, the polynomial $\Sigma^{(+)}$ has four strictly positive roots, $\{1,1/\sqrt{2},(\sqrt{5}\pm 1)/2\}$, and it turns out that there are three ranges of stability for the background solution,
\begin{equation}\label{eq_stab_range_d3}
    B_0 \in \bigg] 0 ; \frac{\sqrt{5}-1}{2}\bigg[\,,
    \qquad 
    B_0 \in \bigg] \frac{1}{\sqrt{2}} ; \sim 0,738\bigg[
    \qquad\text{and}\qquad
    B_0 \in \bigg] 1 ; \frac{\sqrt{5}+1}{2}\bigg[\,.
\end{equation}

When $d = 4$, it simply comes
\begin{equation}
    \mu_+^{(+),d=4} = \frac{3}{2}\Bigg[\sqrt{\frac{9\,B_0^4-16\,B_0^2-20}{9\,B_0^2(B_0^2-4)}}-1\Bigg]\,.
\end{equation}
The eigenvalue thus has a negative real part in the range $B_0 \in ]1;2[$.\\

Therefore, when $\kappa = 1$ and $d \geq 3$, there is always at least one range of parameters on which the compactified solution is an attractor.

\subsection{No-ghost conditions for the tensor modes}\label{sec_noghost}

Let us now investigate the fate of the tensorial modes in the regions where the background solution is an attractor. 
We restrict the study to the transverse and traceless (TT) perturbation modes of each sub-manifold.
Introducing $\hat{\partial}_i$ and $\eth_A$ the covariant derivatives compatible respectively with $\hat{\delta}_{ij}$ and $\omega_{AB}$, and operating the indices of sub-manifolds by the corresponding spatial metric, the line element~\eqref{eq_lineelement_cosmobckg} is perturbed as
\begin{equation}\label{eq_lineelement_pert}
    \dd s^2  =-\dd t^2 + a^2(t)\bigg[\hat{\delta}_{ij}+h_{ij}\bigg]\dd x^i \dd x^j + b^2(t) \,\bigg[\omega_{AB}+H_{AB}\bigg]\,\dd x^A \,\dd x^B\,,
\end{equation}
where the perturbation obeys the TT conditions
\begin{equation}\label{eq_TT_cond}
    \hat{\delta}^{ij}h_{ij} = \hat{\partial}^jh_{ij} = 0 = \eth^B H_{AB} = \omega^{AB}H_{AB}\,.
\end{equation}
Naturally, the perturbation $h_{ij}$ behaves as a scalar regarding the $\eth_A$ derivative (and reversely, $H_{AB}$ behaves as a scalar regarding the $\hp_i$ derivative).\\ 

As presented in detail in App.~\ref{app_pert_scenario1}, those two perturbations naturally decouple in quadratic order in the action.
%ADF
We are splitting the perturbation fields according to their behavior under three-dimensional (3D) rotations, subspace of our four-dimensional world, and according to the rotations in the extra dimensions. Assuming the background is homogeneous and isotropic also in the extra dimensions, we will have invariance under rotations about the $d$ dimensional spatial extra-dimensions. Therefore we can distinguish all the perturbations of the metric as 3D-scalars, d-scalars, 3D-vectors, $d$-vectors, etc. For instance, we could have 3D scalars which are $d$-tensor models, and we have called these modes $H_{AB}$. On the contrary, we could have $d$-scalars (according to $d$-dimensional rotations) which behave as 3D tensors (according to 3D rotations). These modes are called $h_{ij}$. At a linear level, it is easy to see that these modes will decouple from any other propagating mode; in particular, they will not couple with each other. This implies we can study their dynamics independently of what happens for all the other modes. This considerably simplifies the analysis. Although this stability study will not be complete, we can still give the necessary conditions for the stability of the background regarding the 3D tensor and $d$-tensor modes.
%ADF

Injecting the compactified solution~\eqref{eq_compactification_ansatz} and using the dimensionless variables~\eqref{eq_adim_var}, it comes
\begin{equation}\label{eq_d2S_tens}
    \delta^{(2)}\mathcal{S} = \frac{\Mpl^2}{8}\int\!\!\dd t\,\dd^3x\,\dd^d x\,\sqrt{\omega}\,a^3b_0^d\,H_0^2\,\Bigg\lbrace
    \mathcal{L}^{(2)}_\text{phys}\big[h_{ij}\big]
    + \mathcal{L}^{(2)}_\text{extr}\big[H_{AB}\big]
    \Bigg\rbrace\,,
\end{equation}
where $\sqrt{\omega}$ is the squared root of the determinant of the angular metric associated with the extra-dimensional sub-manifold $\mathcal{M}_d$, and 
\begin{subequations}\label{eq_d2S_tens_detail}
    \begin{align}
        & 
        \mathcal{L}^{(2)}_\text{phys}  =
        \frac{\mathcal{K}}{H_0^2}\Bigg[\left(\partial_t h_{ij}\right)^2 - \left(\frac{\hp_k h_{ij}}{a}\right)^2- c_\text{extr}^2\left(\frac{\partial_A h_{ij}}{b_0}\right)^2\Bigg] - \mathcal{M}^2\,h_{ij}^2\,,\\
        &
        \mathcal{L}^{(2)}_\text{extr} =
        \frac{\widetilde{\mathcal{K}}}{H_0^2}\Bigg[\left(\partial_t H_{AB}\right)^2 - \left(\frac{\partial_i H_{AB}}{a}\right)^2 - \tilde{c}_\text{extr}^2\left(\frac{\eth_C H_{AB}}{b_0}\right)^2\Bigg] - \widetilde{\mathcal{M}}^2\,H_{AB}^2\,.
    \end{align}
\end{subequations}
We recall that $h_{ij}$ behaves as a scalar for the covariant derivative $\eth_A$ (and, respectively, $H_{AB}$ behaves as a scalar for the covariant derivative $\hp_i$), hence the presence of partial derivatives in the quadratic Lagrangians.
The constants entering this quadratic action read
\begin{subequations}
\begin{align}
    & \Bigg. \label{eq_noghost_phys}
    \mathcal{K} = 
    1 +2d(d-1)\frac{\kappa \, \beta}{B_0^2}
    \,,\\
    & \Bigg.
    c_\text{extr}^2 = 1 + 4\beta\,\frac{1-(d-1)\frac{\kappa}{B_0^2}}{1+2d(d-1)\frac{\kappa \, \beta}{B_0^2}}\,,\\
    & \Bigg.
    \mathcal{M}^2 =
    8-2\lambda-80\beta+d(d-1)\big(1-32\,\beta\big)\frac{\kappa}{B_0^2}+d(d-1)(d-2)(d-3)\frac{\kappa^2\beta}{B_0^4}
    \,,\\
    & \Bigg. \label{eq_noghost_extr}
    \widetilde{\mathcal{K}} =
    1 + 12\,\beta+2(d-2)(d-3)\frac{\kappa \, \beta}{B_0^2}
    \,,\\
    & \Bigg.
    \tilde{c}_\text{extr}^2 = 1 + 4\beta\,\frac{3-(d-3)\frac{\kappa}{B_0^2}}{1+12\,\beta+2(d-2)(d-3)\frac{\kappa \, \beta}{B_0^2}}
    \,,\\
    & \Bigg.
    \widetilde{\mathcal{M}}^2 =12-2\lambda+24\,\beta +\big(d^2-3d+4\big)\big(1+24\,\beta\big)\frac{\kappa}{B_0^2} + (d-3)(d-4)\big(d^2-3d+6\big)\frac{\kappa^2\beta}{B_0^4}
    \,,
\end{align}
\end{subequations}
where we have not injected the background solution~\eqref{eq_lambda_beta_sol} for the sake of clarity.\\

To demonstrate that there is no range of stability for the compactified solution, it will be sufficient to focus on the two ``no-ghost'' conditions, $\mathcal{K}$ and $\widetilde{\mathcal{K}}$, that should be simultaneously positive to avoid a ghostly gravitational mode.

\subsubsection{Case $d = 3$}

As it has been shown in Sec.~\ref{sec_background_stab_d2}, in the case $d = 2$, the background is never stable. 
Hence we start by considering $d=3$, for which the background solution~\eqref{eq_lambda_beta_sol} reads
\begin{equation}
    \beta = \frac{3B_0^2-2\kappa}{12(\kappa-B_0^2)}\,,
    \qquad
    \lambda =9-\frac{6B_0^4-3}{B_0^2(B_0^2-\kappa)}\,.
\end{equation}
The no-ghost condition thus read
\begin{equation}
    \mathcal{K} = 
    \frac{B_0^4-4\kappa\,B_0^2+2}{B_0^2(B_0^2-\kappa)}
    \qquad\text{and}\qquad
    \widetilde{\mathcal{K}} =
    -\frac{2\,B_0^2-\kappa}{B_0^2-\kappa}\,.
\end{equation}
When $\kappa= -1$, the ratio of the two is always negative, which means that one of the two conditions is never fulfilled.
When $\kappa= 1$, they are simultaneously positive only the range $B_0 \in ]\sqrt{2-\sqrt{2}};1[$, which has no overlap with the three ranges of background stability listed in Eq.~\eqref{eq_stab_range_d3}. 
Hence the compactification scenario is not stable with three extra dimensions.

\subsubsection{Case $d = 4$}

In the case of four extra dimensions, treated in more detail in Sec.~\ref{sec_example}, the background solution is
\begin{equation}
    \beta = -\frac{(B_0^2-\kappa)B_0^2}{4(B_0^4-1)}\,,
    \qquad
    \lambda =- \frac{3(B_0^2-\kappa)(B_0^4-3\kappa\,B_0^2+1)}{B_0^2(B_0^4-1)}\,.
\end{equation}
The no-ghost condition thus read
\begin{equation}
    \mathcal{K} = 
    \frac{(B_0^2-5\kappa)(B_0^2-\kappa)}{B_0^4-1}
    \qquad\text{and}\qquad
    \widetilde{\mathcal{K}} =
    -\frac{2\,B_0^2(B_0^2-\kappa)}{B_0^4-1}\,.
\end{equation}
When $\kappa = -1$, they have opposite signs, and when $\kappa = 1$, $\widetilde{\mathcal{K}} = -2B_0^2/(B_0^2+1)$ is always negative. Thus, the compactification scenario is not stable with four extra dimensions.

\subsubsection{Case $d \geq 5$}

In the case of more than 4 extra dimensions, the no-ghost conditions read
\begin{subequations}
\begin{align}
    &
    \mathcal{K} = 
    \frac{6B_0^4-12(d-1)\kappa\,B_0^2+2(2d-3)(d-1)}{6B_0^4+3(d-1)(d-4)\kappa\,B_0^2-(d-1)(d-2)(d-3)}\,,\\
    &
    \widetilde{\mathcal{K}} =
    - \frac{6B_0^2[2B_0^2-(d-2)\kappa]}{6B_0^4+3(d-1)(d-4)\kappa\,B_0^2-(d-1)(d-2)(d-3)}\,.
\end{align}
\end{subequations}

When $\kappa = -1$, the ratio
\begin{equation}
     \frac{\mathcal{K}}{\hat{\mathcal{K}}} = -\frac{3\,B_0^4+6(d-1)\,B_0^2+(d-1)(2d-3)}{3B_0^2[2\,B_0^2+(d-2)]}\,,
\end{equation}
is obviously negative (as $d>2$).\\

When $\kappa = 1$, $\mathcal{K}$ and $\hat{\mathcal{K}}$ are simultaneously positive in the range
\begin{equation}
    B_0 \in \Bigg] \frac{1}{2}\sqrt{-(d-1)(d-4)+\sqrt{\frac{d(d-1)(3d^2-19d+32)}{3}}} ; \sqrt{d-1-\sqrt{\frac{d(d-1)}{3}}} \Bigg[\,,
\end{equation}
which upper limit is smaller than $B_0^{(\Pi,1)}$, see Eq.~\eqref{eq_root_B0Pi1}. 
Thus, there is no overlap between the range of stability of the tensorial perturbations and the background ones and, even for more than four extra dimensions, the compactification scenario is not viable.

\section{A concrete example: the case of four extra dimensions}\label{sec_example}

We rediscuss the previous analysis in terms of an explicit example by fixing the dimensions of four extra dimensions, namely $d=4$. On top of that, we will also make use of explicit eigenfunction expressions, as to give an independent check of our previous results, and possibly to give a different point of view in describing still the same instability issue. In this section, we take the three-dimensional metric to be Euclidean $\dd s^2_{3D} = \delta_{ij}\dd x^i \dd x^j$.\\

We have already discussed the existence of two sets of tensor modes, the tensor modes in three dimensions and the ones in $d=4$. However, the former ones behave as tensors concerning three-dimensional rotations and scalars according to $d=4$ rotations, and vice-versa, the latter behave as scalars according to three-dimensional rotations and as tensors concerning $d=4$ rotations. It is then clear that both these modes will decouple from any other modes. The three-dimensional tensor modes satisfy $h_{ij}=h_{ji}$, $\delta^{ij} h_{ij}=0$, and $\delta^{ij} \partial_i h_{jk} = 0$, which leave only two independent components. Given the symmetries of the background, since we are interested in linear theory, all Fourier modes will decouple from each other, allowing us to study the general properties of their propagation in a minimal simplified manner. In particular, we can choose those modes that propagate in one three-dimensional spatial and one $d=4$ spatial dimension.
The $d=4$ tensor modes are defined such that $H_{AB}=H_{BA}$, $\omega^{AB} H_{AB}=0$, and $\omega^{AB} \eth_A H_{BC} = 0$, leaving five independent polarizations.\\

One can then assume propagation in the $x$-dimension and in the extra-dimensional radial dimension, say $r$.
Hence, we can choose $h_{1i}=0$ whereas $h_{ij}=\epsilon_{ij}\sum_{\lambda =1,2}h^\lambda_{ij}(t,x,r)$, for which $\epsilon_{1j}=0$, $\epsilon_{ij}=\epsilon_{ji}$, $\epsilon_{22}=-\epsilon_{33}$. 
We can choose $h^\lambda$ to be eigenfunction of the standard Laplacian $\delta^{ij} \partial_i\partial_j h^\lambda = -q^2 h^\lambda$, and to be an eigenfunction of the $d=4$ Laplacian, $\omega^{AB} \eth_A \eth_B h^\lambda = -k^2 h^\lambda $, where we recall that $\eth_A$ is the covariant derivative compatible with $\omega_{AB}$.\footnote{An $r$-dependent eigenfunction $\chi$ satisfies the relation $\omega^{AB}\eth_{A}\eth_{B}\chi=(1-\kappa\,r^{2})\,\chi''+\frac{3-4\,\kappa\,r^{2}}{r}\,\chi'=-k^{2}\,\chi$.} Out of the action for the perturbations we find the equation of motion as $\frac{\delta S}{\delta h^\lambda}=E_\lambda$, and up to the background value of $\sqrt{-g}$, we find
\begin{align}
    E_\lambda=-\frac{\left(B_{0}^{2}-5\kappa\right) \ddot{h}_{\lambda }}{2 \left(B_{0}^{2}+\kappa\right) H_0^{4}}
    +\frac{3 \left(5\kappa-B_{0}^{2}\right) \dot{h}_{\lambda}}{2 H_0^{3} \left(B_{0}^{2}+\kappa\right)}
    -\frac{\left(B_{0}^{2}-5\kappa\right) q^2  h_{\lambda} }{2 H_0^{4} \mathit{a}^{2} \left(B_{0}^{2}+\kappa\right)}
    +\frac{\kappa \,k^{2} h_{\lambda}}{B_{0}^{2} H_0^{2} \left(B_{0}^{2}+\kappa\right)}=0\,.
\end{align}

The no-ghost condition to be imposed is
\begin{equation}
    \frac{B_0^2-5\kappa}{B_0^2+\kappa}>0\,,
    \qquad\text{hence}\qquad
   \{B_0^2 >5 \;{\rm  and}\; \kappa=1\}
    \quad {\rm  or}\quad  \{B_0^2 >1 \;{\rm  and} \;\kappa=-1\}\,.
\end{equation}
and the positivity of the speeds of propagation reads
\begin{equation}
    c_{3D}^2 = 1\,,\qquad c_{d=4}^2 = \frac{2}{5-\kappa\, B_0^2}>0\,.
\end{equation}
The last relation, describing the speed in the extra dimensions, is trivially satisfied for $\kappa=-1$. For $\kappa=1$, we require that $B_0^2<5$, which is not compatible with the no-ghost condition. Therefore, the three-dimensional tensor modes either are ghosts or the background is unstable. The background solution is stable if $1<B_0<2$ so that the three-dimensional tensor modes are ghost degrees of freedom. But even without the stability of the background, this shows that there will always be some issues with the three-dimensional tensor models if $\kappa=1$. Finally, we notice that the three-dimensional tensor modes do not acquire any mass in the extra dimensions.\\

Let us focus our attention on the $d=4$ tensor modes. We can now find a suitable choice of eigenfunctions. For instance, although the following holds for the five independent polarizations, let us focus on only one of them for instance $H_{A=2, B=4}=H^{(1)}(t,x,r)(1-w^2)/(1-z^2)^{3/2}$, where $w$ and $z$ represent the cosines of the two polar angles parametrizing the 3-sphere.
The dynamics of the $H^{(1)}$ field polarization decouples from the other and we find its equation of motion giving
\begin{align}
    E^{(1)}=-\frac{B_{0}^{2} \ddot{H}^{(1)}}{-\kappa-B_{0}^{2}}
    +\frac{3 B_{0}^{2} H_{0} \dot{H}^{(1)}}{B_{0}^{2}+\kappa}
    +\left(\frac{\left(5 B_{0}^{2} -\kappa\right)H_0^2 k^{2}}{2 B_{0}^{2}(B_{0}^{2}+\kappa) }+\frac{ B_{0}^{2} q^{2}}{\left(B_{0}^{2}+\kappa\right) a^2}
    +\frac{H_{0}^{2} \left(5\kappa B_{0}^{2}-1\right)}{B_{0}^{2}(B_{0}^{2}+\kappa)}\right) H^{(1)}=0\,,
\end{align}
out of which we can deduce the no-ghost condition
\begin{equation}
    -\frac{2B_0^2}{\kappa+B_0^2}>0\,,
\end{equation}
which is never satisfied for $\kappa=1$, leading always to $d=4$ tensor ghost degrees of freedom. Instead for $\kappa=-1$, we require $0<B_0^2<1$. We can now read the speeds of propagation for the $d=4$ tensor modes, as
\begin{align}
    c_{3D}^2=1\,,\qquad c_{d=4}^2=\frac{5B_0^2-\kappa}{2B_0^2}\,.
\end{align}
For $\kappa=-1$, this quantity is always positive. The problem with $\kappa=-1$  is that either the three-dimensional tensor or the $d=4$ tensor modes are ghosts. This happens because the  $d=4$ tensors are not ghost if $0<B_0^2<1$, however, the three-dimensional tensor modes in this range are ghost degrees of freedom. Vice-versa, if $B_0^2>1$ the three-dimensional tensor are not ghost degrees of freedom but the $d=4$ ones are. The same equations of motion and constraints hold for the remaining four polarization eigenfunctions.
We have confirmed, taking a different path, that the explicit eigenfunction method gives the same constraints as in the general abstract formalism, at least for $d=4$.

\section{Changing the dynamics entails stable compactification scenarios}\label{sec_change}

As proven in Sec.~\ref{sec_instab}, the original compactification scenario~\eqref{eq_compactification_ansatz} does not lead to viable configurations.
It is thus tempting to seek minimal changes that would allow stable compactified solutions.\\

To do so, we relax the $b \to b_0$ hypothesis, without changing the dynamics of the physical sector ($H = H_0$ will still be taken as a positive constant).
Naturally, time parametrization invariance still allows us to take $N = 1$ without loss of generality and we will use the dimensionless variables $\{\beta,\lambda\}$ defined in Eq.~\eqref{eq_adim_var}. 

\subsection{An interesting solution in the case of a flat extra-dimensional sub-manifold}

The equations of motion~\eqref{eq_eom_bckg} are second-order in $b$. 
But it is easy to note that each occurrence of $b$ is associated with $\kappa$. 
Hence, taking a flat extra-dimensional sub-manifold drastically simplifies the problem, as it reduces the order of the equations of motion.
Taking $\kappa=0$, the constraint~\eqref{eq_eom_bckg_N} reads indeed
\begin{equation}
\begin{aligned}
    \frac{\mathcal{C}}{H_0^2} = \
    &
    \frac{d(d-1)(d-2)(d-3)\beta}{2}\bigg(\frac{L}{H_0}\bigg)^4
    + 6d(d-1)(d-2)\beta\bigg(\frac{L}{H_0}\bigg)^3\\
    & \qquad\quad
    +\frac{d(d-1)(1+36\beta)}{2}\bigg(\frac{L}{H_0}\bigg)^2
    + 3d(1+4\beta)\,\frac{L}{H_0}
    +3-\lambda = 0\,.
\end{aligned}
\end{equation}
Being a polynomial in $L/H_0$, this constraint is solved either by taking $L/H_0 = \mathcal{X}_0$ constant (but different from zero, otherwise we encounter the problem discussed in Sec.~\ref{sec:flatEDfinetuning}) or by canceling each coefficient.
The second solution leads to the strongly tuned, trivial solution $\{d = 1,\beta = -1/4,\lambda=3\}$, and thus will not be considered.\\

Considering $L = H_0\,\mathcal{X}_0$ constant, the equations of motion are solved either by
\begin{equation}
    \mathcal{X}_0 = 1 \qquad\text{and} \qquad \lambda = \frac{(d+2)(d+3)}{2}\bigg[1 + d(d+1)\beta\bigg]\,,
\end{equation}
or by
\begin{subequations}\label{eq_bckg_sol_k0}
    \begin{align}
    & \beta = \frac{-1}{4+8(d-1)\mathcal{X}_0+2(d-1)(d-2)\mathcal{X}_0^2}\,,\\
    & \lambda = \frac{24+48(d-1)\mathcal{X}_0+4(d-1)(7d-6)\mathcal{X}_0^2+8d(d-1)^2\mathcal{X}_0^3+(d+1)d(d-1)(d-2)\mathcal{X}_0^4}{8+16(d-1)\mathcal{X}_0+4(d-1)(d-2)\mathcal{X}_0^2}\,.
    \end{align}
\end{subequations}
The first solution implies an expanding behaviour, $b = b_0 \, e^{H_0 t}$, which is not compatible with the sought compactification scenario.\\

As for the second solution, it does not suffer from fine-tuning issues and is indeed a compactified solution as long as $\mathcal{X}_0 < 0$.
The compactification scenario that will be studied hereafter is thus described by
\begin{equation}\label{eq_compactification_ansatz_k0}
    N(t) \to 1\,,
    \qquad
    H(t) \to H_0\,,
    \qquad
    L(t) \to H_0\,\mathcal{X}_0
    \qquad\text{and}\qquad
    \kappa = 0\,.
\end{equation}

\subsubsection{Attractor behavior of this solution}

Let us first study the attractor behavior of the previously found solution.
In the spirit of what was done in Sec.~\ref{sec_attractor}, one can use
\begin{equation}
    H = H_0\big(1 + \delta H\big)\,,
    \qquad
    L = H_0\big(\mathcal{X}_0 + \delta \mathcal{X}\big)\,.
\end{equation}
As the dynamical system~\eqref{eq_eom_bckg} is redundant, the constraint equation can be eliminated so that it comes
\begin{equation}
    H_0\,\begin{pmatrix}
        \delta \dot{H} \\
        \delta \dot{\mathcal{X}}
    \end{pmatrix}
    = -\big(3 + d\,\mathcal{X}_0\big)\,
    \begin{pmatrix}
        \delta H \\
        \delta \mathcal{X}
    \end{pmatrix}
    + \mathcal{O}(2)\,,
\end{equation}
which is obviously stable as long as $\mathcal{X}_0 > -3/d$. 
Hence, we have restricted the allowed parameter space for $\mathcal{X}_0$ to $-3/d<\mathcal{X}_0<0$.

\subsubsection{No-ghost condition}

As for the fate of the tensorial modes, using the machinery developed in Sec.~\ref{sec_noghost} and the explicit computation performed in App.~\ref{app_pert_scenario2}, the two perturbation decouple as
\begin{equation}\label{eq_d2S_tens_k0}
    \delta^{(2)}\mathcal{S} = \frac{\Mpl^2}{8}\int\!\!\dd t\,\dd^3x\,\dd^dx\,a^3b^d\,H_0^2\,\Bigg\lbrace
    \mathcal{L}^{(2)}_\text{phys}\big[h_{ij}\big]
    + \mathcal{L}^{(2)}_\text{extr}\big[H_{AB}\big]
    \Bigg\rbrace\,,
\end{equation}
where
\begin{subequations}\label{eq_d2S_tens_detail_k0}
    \begin{align}
        & 
        \mathcal{L}^{(2)}_\text{phys}  =
        \frac{\mathcal{K}}{H_0^2}\Bigg[\left(\partial_t h_{ij}\right)^2 - c_\text{phys}^2\left(\frac{\hp_k h_{ij}}{a}\right)^2- c_\text{extr}^2\left(\frac{\partial_A h_{ij}}{b}\right)^2\Bigg] - \mathcal{M}^2\,h_{ij}^2\,,\\
        &
        \mathcal{L}^{(2)}_\text{extr} =
        \frac{\widetilde{\mathcal{K}}}{H_0^2}\Bigg[\left(\partial_t H_{AB}\right)^2 - \tilde{c}_\text{phys}^2\left(\frac{\partial_i H_{AB}}{a}\right)^2 - \tilde{c}_\text{extr}^2\left(\frac{\eth_C H_{AB}}{b}\right)^2\Bigg] - \widetilde{\mathcal{M}}^2\,H_{AB}^2\,.
    \end{align}
\end{subequations}
The expressions of the constant entering those quadratic actions are presented in Eq.~\eqref{eq_constant_k0}.\\

The no-ghost conditions, 
\begin{equation}
    \mathcal{K} = 
    \frac{2(1-\mathcal{X}_0)[1+(d-1)\mathcal{X}_0]}{2+(d-1)\mathcal{X}_0[4+(d-2)\mathcal{X}_0]}
    \qquad\text{and}\qquad
    \widetilde{\mathcal{K}} =
   -\frac{2(1-\mathcal{X}_0)[2+(d-2)\mathcal{X}_0]}{2+(d-1)\mathcal{X}_0[4+(d-2)\mathcal{X}_0]}
    \,,
\end{equation}
are simultaneously positive when\footnote{We recall that $\mathcal{X}_0$ has to be strictly negative to achieve a compactification scenario.}
\begin{equation}\label{eq_range_XO_k0}
    \mathcal{X}_0 \in \bigg] \frac{-2}{d-2} ; \frac{-1}{d-1}\bigg[\,.
\end{equation}
When $d= 2$, the range of stability spans every negative number up to $-1/2$.
This range is naturally compatible with the attractor condition $\mathcal{X}_0 > -3/d$.\\

However, and contrarily to the original compactification scenario, the speed of propagation of gravitational waves is not unity in the physical sector, as
\begin{equation}
    c_\text{phys}^2 = 1  + \frac{d\,\mathcal{X}_0}{1+(d-1)\,\mathcal{X}_0}\,.
\end{equation}
Therefore, in order to be consistent with the bounds $\vert c^2_\text{GW}/c^2_\text{light}-1 \vert \lesssim 10^{-15}$~\cite{LIGOScientific:2017zic}, we need to impose
\begin{equation}
    \vert \mathcal{X}_0 \vert \lesssim \frac{10^{-15}}{d}\,,
\end{equation}
which is not compatible with the range~\eqref{eq_range_XO_k0} where both no-ghost conditions are verified. In fact, for such values of $|\mathcal{X}_0|$, we can see that $\tilde{\mathcal{K}}\approx-2$.\\

The compactification scenario~\eqref{eq_compactification_ansatz_k0} is interesting as, for any number of dimensions, it seems to enjoy a region of parameter space where it is stable. 
However its phenomenological implications are not compatible with observational bounds, and the model should be discarded.

\subsection{The case of non-flat extra-dimensional sub-manifolds}

In the case where $\kappa = \pm 1$, the coefficient of $b^{-4}$ in the constraint equation~\eqref{eq_eom_bckg_N} is proportional to $(d-2)(d-3)$.
This implies that either $d$ is 2 or 3 or other contributions of the equations of motion with similar time dependencies ``compensate'' this factor.

\subsubsection{Case $d = 2$}

In the case $d= 2$, the combination
\begin{equation}
    \frac{1+12\beta}{2}\frac{\mathcal{E}_a-\mathcal{E}_b}{H_0^2} + \bigg(1+4\beta + 8\beta\,\frac{L}{H_0}\bigg)\frac{\mathcal{C}}{H_0^2} = \bigg(1+4\beta + 8\beta\,\frac{L}{H_0}\bigg)\bigg(6+12\beta-\lambda+4(1+12\beta)\frac{L}{H_0}\bigg)=0
\end{equation}
imposes that $L$ is constant.
Solving further the system, it comes $L_0 = H_0$, $\beta= -1/12$, and $\lambda = 5$, which is a strongly tuned expanding solution, hence of no interest for this work.

\subsubsection{Case $d = 3$}

In the case $d = 3$, one can first solve for $\lambda$ the constraint, Eq.~\eqref{eq_eom_bckg_N}, and inject it in the equation of motion for $a$, yielding
\begin{equation}
    \mathcal{E}_a = 12 \bigg( \dot{L} + L^2 - H_0\,L\bigg)\bigg(\beta\,L^2+4\beta\,H_0\,L+ \frac{1+4\beta}{4}\,H_0^2 + \frac{\beta\,\kappa}{b^2}\bigg)=0\,.
\end{equation}
So there are two branches of solution.
The first branch $\dot{L} + L^2 - H_0L = 0$ is solved by
\begin{equation}
    L = \frac{H_0}{1+c_1 e^{-H_0 t}}\,,
    \qquad\text{and}\qquad
    b = c_2\,\big[c_1 + e^{H_0 t}\big]\,,
\end{equation}
which is not a compactified solution.
As for the second branch, one can solve the algebraic sector for $L$ and inject the solution into the remaining equation.
This gives
\begin{equation}
    \mathcal{E}_b = 2\,\bigg(\dot{L}- \frac{\kappa}{b^2}\bigg)\Bigg[1-12\,\beta+ 6\beta\sqrt{\bigg(12\beta-1- \frac{4\kappa\,\beta}{H_0^2\,b^2}\bigg)\beta}\Bigg]=0\,.
\end{equation}
The differential sector of this equation yields solutions that are not compatible with the previously found constraint on $L$.
As for the algebraic sector, it imposes a constant $b$, which has already been treated in Sec.~\ref{sec_instab}.

\subsubsection{Case $d=4$}

To illustrate the behavior of the system for higher than 3 extra dimensions, let us study the $d = 4$ case.
As previously, we suppose a de Sitter expansion for the physical manifold, namely $a\propto \exp(H_0 t)$, and impose $L\neq 0$ and $\kappa\neq0$. On considering an appropriate combination of the lapse $N$-equation of motion and the $a$-equation of motion, we find the following 
\begin{align}
    H_0^2\,\frac{\mathcal{E}_a-\mathcal{E}_N}{48}=
    \bigg(\beta\,L^2+2\beta\,H_0\,L+ \frac{1+4\beta}{12}\,H_0^2 + \frac{\beta\,\kappa}{b^2}\bigg)\bigg( \dot{L} + L^2 - H_0\,L\bigg)=0\,,
\end{align}
so that we have two branches for the solution. Let us formally solve the first, algebraic branch, finding an equation determining $b^2$ as a function of $L$. On replacing this equation for $b^2$ into the $N$-equation of motion, we get another equation that is polynomial in $L$ with constant coefficients, leading, in general, to $L={\rm const}$. But, according to our assumptions, this solution cannot be accepted, as a constant non-zero $L$ would also lead to a constant $b$, leading to inconsistencies (as $L={\dot b}/b$).\\

The other branch, for which $\dot L = H_0L-L^2$, leads again to
\begin{equation}
    L = \frac{H_0}{1+c_1 e^{-H_0 t}}\,,
    \qquad\text{and}\qquad
    b = c_2\,\big[c_1 + e^{H_0 t}\big]\,.
\end{equation}
On substituting these relations in the $a$-equation of motion we obtain a polynomial in $\exp(H_0 t)$ of fourth-order, whose coefficients need all to vanish. But by doing this we overconstrain the system. In any case, $b$ would expand, i.e.\ not realizing a satisfactory compactification scenario.

\subsubsection{Case of higher dimensions}

For more than 4 extra dimensions, we need to ``compensate'' the $b^{-4}(t)$ dependency.
Excepted for the already treated case $b(t) = b_0$, the only way of doing so would be to impose $L \propto b^{-n}$ or $\dot{L} \propto b^{-2}$.
The first case yields $b(t) \propto (1 + t/\tau_0)^\frac{1}{n}$ and the second, $b(t) \propto \sinh[\mu_0(t-\tau_0)]$.
Both cases are not compactification scenarios, and we will not consider them.\\

Therefore it seems that there is no viable solution when relaxing only the hypothesis $b = b_0$.
One should thus carry the full study and seek for the attractor regions of the whole $\{b, H\}$ plane, which is left for future work.

\section{Conclusions}\label{sec_concl}

In more than four dimensions, the Einstein-Gauss-Bonnet framework allows for an interesting compactification scenario in the cosmological context.
Once a de Sitter configuration is taken for the physical (four-dimensional) space, the extra-dimensional sub-manifold wraps up towards a configuration with a constant scale factor.
However, as we proved in this work, such configurations are inherently unstable: while the background may act as an attractor for certain parameter ranges, tensorial perturbations exhibit a ghost-like behavior in these regimes.
The addition of matter has the potential to alter the background configuration and modify its attractor properties. 
By introducing appropriately chosen matter components, it may be possible to reconcile the stability of the background with that of the tensor modes, creating overlapping regions of stability. 
This compelling possibility warrants further exploration and is left as a direction for future research.\\

In this study, we attempted to circumvent the instabilities by relaxing one of the core assumptions of the compactification scenario. Specifically, we allowed the scale factor of the extra-dimensional sub-manifold to vary with time. Under this relaxation, no solutions were found for cases where the extra-dimensional sub-manifold possesses non-zero intrinsic curvature. However, in the case of a flat sub-manifold, stable configurations with an exponentially decreasing scale factor were identified. Unfortunately, these configurations are inconsistent with current observational constraints on the speed of gravitational wave propagation and must therefore be ruled out.\\

To comprehensively explore compactification scenarios within the Einstein-Gauss-Bonnet framework, it is necessary to relax the de Sitter assumption for the physical spacetime and map the attractor regions across the full space of configurations. The stability of gravitational perturbations in these configurations can then be systematically investigated. This promising avenue of research remains an exciting prospect for future work.

\acknowledgments

The authors would like to express their gratitude to A.~V.~Toporensky for illuminating discussions and questions, as well as to A.~Giacomini for an interesting exchange at an early stage of the project.
The work of ADF was supported by the JSPS Grants-in-Aid for Scientific Research No.~20K03969.

%=======================================================
\appendix

\section{Lenghty expressions}\label{app_lengthy}

This appendix collects expressions that are too long to be decently presented in the text.

\subsection{Cosmological background equations of motion}\label{app_eom_bckg}

Injecting the metric~\eqref{eq_lineelement_cosmobckg} in the action~\eqref{eq_action_base} and varying it \emph{wrt.} $\{N,a,b\}$ gives the background equations of motion
\begin{subequations}\label{eq_eom_bckg}
\begin{align}
    \mathcal{E}_a
    & \Bigg. \nn
    = -\Lambda + \frac{d(d-1)\,\kappa}{2\,b^2} + 3H^2+2d\,HL+\frac{d(d+1)}{2}\,L^2+\frac{2\dot{H}}{N}+\frac{d\dot{L}}{N}\\
    & \Bigg. \quad \nn
    + \alpha \Bigg[
    \frac{d(d-1)(d-2)(d-3)\,\kappa^2}{2\,b^4}
    + \frac{d(d-1)\,\kappa}{b^2}\bigg(6H^2+4(d-2)HL+(d-1)(d-2)L^2\bigg)\\
    & \Bigg. \qquad\qquad \nn
    + 8d\,H^3L
    + 2d(5d-3)H^2L^2
    + 4d^2(d-1)HL^3
    + \frac{(d+1)d(d-1)(d-2)}{2}\,L^4\\
    & \Bigg. \qquad\qquad \nn
    + \frac{4d(d-1)\,\kappa}{b^2}\bigg(\frac{\dot{H}}{N}+ \frac{d-2}{2}\frac{\dot{L}}{N}\bigg)
    +8d\bigg(HL+\frac{(d-1)L^2}{2}\bigg)\frac{\dot{H}}{N}\\
    & \Bigg. \qquad\qquad
    +4d\bigg(H^2+ 2(d-1)HL+\frac{(d-1)(d-2)L^2}{2}\bigg)\frac{\dot{L}}{N}
    \Bigg]
    \,,
    \label{eq_eom_bckg_a}\\
    \mathcal{E}_b
    & \Bigg. \nn
    = -\Lambda + \frac{(d-1)(d-2)\,\kappa}{2\,b^2} + 6H^2+3(d-1)\,HL+\frac{d(d-1)}{2}\,L^2+\frac{3\dot{H}}{N}+\frac{(d-1)\dot{L}}{N}\\
    & \Bigg. \quad \nn
    + \alpha \Bigg[
    \frac{(d-1)(d-2)(d-3)(d-4)\,\kappa^2}{2\,b^4}
    + \frac{(d-1)(d-2)\,\kappa}{b^2}\bigg(12H^2+6(d-3)HL+(d-2)(d-3)L^2\bigg)\\
    & \Bigg. \qquad\qquad \nn
    + 12H^4
    + 36(d-1)\,H^3L
    + 12(d-1)(2d-3)H^2L^2
    + 6(d-1)^2(d-2)HL^3\\
    & \Bigg. \qquad\qquad \nn
    + \frac{d(d-1)(d-2)(d-3)}{2}\,L^4
    + \frac{6(d-1)(d-2)\,\kappa}{b^2}\bigg(\frac{\dot{H}}{N}+ \frac{d-3}{3}\frac{\dot{L}}{N}\bigg)\\
    & \Bigg. \qquad\qquad \nn
    +12\bigg(H^2+2(d-1)HL+\frac{(d-1)(d-2)L^2}{2}\bigg)\frac{\dot{H}}{N}\\
    & \Bigg. \qquad\qquad
    +12(d-1)\bigg(H^2+ (d-2)HL+\frac{(d-2)(d-3)L^2}{6}\bigg)\frac{\dot{L}}{N}
    \Bigg]
    \,,
    \label{eq_eom_bckg_b}\\
    \mathcal{C}
    & \Bigg. \nn 
    = -\Lambda + \frac{d(d-1)\,\kappa}{2\,b^2} + 3H^2+3d\,HL+\frac{d(d-1)}{2}\,L^2\\
    & \Bigg. \quad \nn
    + \alpha \Bigg[
    \frac{d(d-1)(d-2)(d-3)\,\kappa^2}{2\,b^4}
    + \frac{d(d-1)\,\kappa}{b^2}\bigg(6H^2+6(d-2)HL+(d-2)(d-3)L^2\bigg)\\
    & \Bigg. \qquad\qquad 
    + 12d\,H^3L
    + 18d(d-1)H^2L^2
    + 6d(d-1)(d-2)HL^3
    + \frac{d(d-1)(d-2)(d-3)}{2}\,L^4
    \Bigg]
    \,.
    \label{eq_eom_bckg_N}
\end{align}
\end{subequations}
Up to a factor 2 in the definition of $\Lambda$ and taking $N(t) = 1$, those are naturally equivalent to, \emph{e.g.}, Eqs.~(3)-(5) of~\cite{Chirkov:2019qbe}.

\subsection{Perturbations of the background solution}\label{app_pert_bckg}

Transforming it to a first-order system by defining $u \equiv \dot{b}$ and setting the lapse function to unity, the dynamical system~\eqref{eq_eom_bckg} can be linearly perturbed as in Eq.~\eqref{eq_pert_bckg_eqlin}, where 
\begin{equation}\label{eq_M0kappa}
    M_0^\kappa = 
    \begin{pmatrix}
        0
        &
        0
        &
        1
        \\
        -\frac{d(d-1)\,M_{21}}{3\,B_0^2\, \Delta_0}
        &
        -\frac{2\,M_{22}}{\Delta_0}
        &
        -\frac{d(d-1)\,B_0\,M_{23}}{\Delta_0}
        \\
        \frac{M_{31}}{3\,B_0^2\, \Delta_0}
        &
        -\frac{2\,B_0\,M_{32}}{\Delta_0}
        &
        -\frac{M_{33}}{\Delta_0}
    \end{pmatrix}\,,
\end{equation}
where
\begin{subequations}
    \begin{align}
    \Delta_0 =\
    & \nn
    12\,B_0^8 - 6(5d-6)\,\kappa\,B_0^6+12(d-1)(2d-3)\,B_0^4\\
    &
    -2(d-1)(d-2)(4d-3)\,\kappa\,B_0^2+d(d-1)(d-2)^2\,,\\
    M_{21} =\
    & \nn
    36\kappa\,B_0^8-66(d-2)B_0^6+42(d-2)^2\kappa\,B_0^4\\
    &
    +(d-2)(11d^2-44d-42)B_0^2+(d-1)(d-2)^2(d-3)\kappa\,,\\
    M_{22} = \
    & \nn
    18\,B_0^8-9(5d-6)\kappa\,B_0^6+2(d-1)(19d-27)B_0^4\\
    &
    -(d-1)(d-2)(14d-9)\kappa\, B_0^2+2d(d-1)(d-2)^2\,,\\
    M_{23} = \ 
    & 
    4\kappa\,B_0^4-4(d-2)B_0^2+(d-2)^2\kappa\,,\\
    M_{31} = \ 
    & \nn
    90(d-2)\kappa\,B_0^8-18(11d^2-40d+32)B_0^6+6(d-1)(d-2)(23d-54)\kappa\,B_0^4\\
    &
    -3(d-1)(d-2)(13d^2-52d+48)B_0^2+2(d-1)^2(d-2)(d-3)(2d-3)\kappa\,,\\
    M_{32} = \ 
    &
    6\kappa\,B_0^6-3(5d-6)B_0^4+2(d-1)(5d-9)\kappa\,B_0^2-(d-1)(d-2)(2d-3)\,,\\
    M_{33} = \ 
    & \nn
    36B_0^8 -18(5d-6)\kappa\,B_0^6-4(d-1)(17d-27)B_0^4\\
    &
    -2(d-1)(d-2)(10d-9)\kappa\,B_0^2+2d(d-1)(d-2)^2\,.
    \end{align}
\end{subequations}
We recall that we use the compactification scenario~\eqref{eq_compactification_ansatz}, the background solution~\eqref{eq_lambda_beta_sol} and the dimensionless variables~\eqref{eq_adim_var}.

\subsection{Attractor behaviour of the $\kappa = 1$ and $d \geq 5$ cases}\label{app_rootB}

When $d \geq 5$ and $\kappa = 1$, the eigenvalue $\mu_+^{(+)}$ has a strictly negative real part on two ranges of $B_0$, 
\begin{equation}
    B_0 \in \bigg] B_0^{(\Pi,1)} ; B_0^{(\Sigma,1)} \bigg[
    \qquad\text{and}\qquad
    B_0 \in \bigg] B_0^{(\Pi,2)} ; B_0^{(\Sigma,2)} \bigg[\,,
\end{equation}
where the upper bounds are the two strictly positive roots of the polynomial $\Sigma^{(+)}$, defined in Eq.~\eqref{eq_def_Pikappa},
\begin{subequations}
    \begin{align}
    &
    B_0^{(\Sigma,1)} = \sqrt{\frac{d}{2}-1}\,,\\
    &
    B_0^{(\Sigma,2)} =  \sqrt{\frac{2(d-1)}{3} + \frac{\aleph_d}{3} + \frac{(d-1)^2}{3\,\aleph_d} }\,,
    \end{align}
\end{subequations}
where we have shortened
\begin{equation}
    \aleph_d \equiv \bigg[1 + \frac{\big(5d^2-15d+6\big)\,d}{4}+\frac{3}{4} \sqrt{d(d-1)^2(d-2)(d-4)(d+2)}\bigg]^{1/3}\,.
\end{equation}
The lower bounds are the two positive roots of the polynomial equation $4\Pi^{(+)}-9\Sigma^{(+)}=0$, namely
\begin{subequations}
    \begin{align}
    & \label{eq_root_B0Pi1}
    B_0^{(\Pi,1)} = \sqrt{\frac{31d^2-118d+96}{60(d-2)}
    + \frac{\sqrt{\Gamma_d^{(1)}}}{2}
    -\frac{\sqrt{\Gamma_d^{(2)}}}{2}}\,,\\
    &
    B_0^{(\Pi,2)} = \sqrt{\frac{31d^2-118d+96}{60(d-2)} 
    + \frac{\sqrt{\Gamma_d^{(1)}}}{2}
    +\frac{\sqrt{\Gamma_d^{(2)}}}{2}}\,,
    \end{align}
\end{subequations}
where we have shortened
\begin{subequations}
    \begin{align}
    \Gamma_d^{(1)} \equiv\ 
    & 
    \frac{\gamma_d}{2}
    + \frac{\beth_d}{3} + \frac{d(d-1)(d-4)^2(d-6)}{225(d-2)\,\beth_d}
    \,,\\
    \Gamma_d^{(2)} \equiv\ \nn
    &  \gamma_d
    -\frac{\beth_d}{3} 
    -\frac{d(d-1)(d-4)^2(d-6)}{225(d-2)\,\beth_d}\\
    &
    + \frac{(d-4)(1531d^5-10250d^4+21940d^3-14040d^2-2880d+3456)}{13500(d-2)^3\sqrt{\frac{\gamma_d}{2}+ \frac{\beth_d}{3}+ \frac{d(d-1)(d-4)^2(d-6)}{225(d-2)\,\beth_d} }}\,,
    \end{align}
\end{subequations}
with the final shortcuts
\begin{subequations}
    \begin{align}
    \gamma_d \equiv \ 
    &
    \frac{(d-4)(121d^3-472d^2+468d-144)}{450(d-2)^2}\,,\\
    \beth_d \equiv
    & 
    \Bigg( \frac{d(d-1)(d-4)^2}{375(d-2)^2} \bigg[
    d^2(d-1)(7d-18)\\
    & \qquad\qquad \nn
    + \sqrt{\frac{d(d-1)(d+2)(2d-3)(73d^4-474d^3+1160d^2-1536d+1152)}{3}} \bigg]\Bigg)^{1/3}\,.   
\end{align}
\end{subequations}
One can naturally check that those roots are accurately ordered, as
\begin{equation}
    0 < B_0^{(\Pi,1)} < B_0^{(\Sigma,1)} < B_0^{(\Pi,2)} < B_0^{(\Sigma,2)}\,.
\end{equation}

\section{Tensorial perturbation of the Einstein-Gauss-Bonnet theory}\label{app_pert_tens}

This appendix presents the details of the computation that yields the quadratic perturbed actions.

\subsection{Generic perturbation}

Let us perturb the line element as
\begin{equation}
    g_{\mu\nu} \,\dd x^\mu \dd x^\nu = \big[\bg_{\mu\nu} + \gh_{\mu\nu}\big]\,\dd x^\mu \dd x^\nu\,,
\end{equation}
Comparing with the definition of the two perturbations $h_{ij}$ and $H_{AB}$, Eq.~\eqref{eq_lineelement_pert}, it is easy to establish the correspondence
\begin{equation}
    \gh_{ij} = a^2(t)\,h_{ij}\,,
    \qquad
    \gh_{AB} = b^2(t)\,H_{AB}\,,    
\end{equation}
all other components of $\gh_{\mu\nu}$ being null.
In the following, bars denote background quantities and $\bn_\mu$ is the covariant derivative compatible with the background metric $\bg_{\mu\nu}$. 
The TT properties of $h_{ij}$ and $H_{AB}$ induce a TT behaviour for $\gh_{\mu\nu}$, namely
\begin{equation}
    \bg^{\mu\nu} \,\gh_{\mu\nu} = \bg^{\mu\nu}\bn_{\mu}\gh_{\nu\rho} = 0\,.
\end{equation}
which can be explicitly checked by using the values of the Christoffel symbols displayed hereafter.\\

Using extensively the TT property of the perturbation as well as Bianchi identities, and performing numerous integrations by parts, the action~\eqref{eq_action_base}
\begin{equation}
    \mathcal{S} = \frac{\Mpl^2}{2}\int\!\!\dd^{d+4}x\sqrt{-g}\bigg\lbrace R - 2\Lambda\bigg\rbrace + \frac{\alpha\,\Mpl^2}{2}\int\!\!\dd^{d+4}x\sqrt{-g}\,\mathcal{G} = \mathcal{S}_\text{EH} + \mathcal{S}_\text{GB}
\end{equation}
is perturbed as
\begin{subequations}
    \begin{align}
    \mathcal{S}_\text{EH} = 
    \frac{\Mpl^2}{2}\int\!\!\dd^{d+4}x\sqrt{-\bg}\Bigg\lbrace
    & \nn 
    \bR -2\Lambda
    - \gh^{\mu\nu} \,\bR_{\mu\nu}
    - \frac{1}{4}\,\big(\bn_\mu \gh_{\nu\rho}\big)^2
    + \frac{1}{2}\,\bR_{\mu\rho\nu\sigma} \,\gh^{\mu\nu}\,\gh^{\rho\sigma} \\
    &
    + \frac{1}{2}\,\bR_{\mu\nu}\, \gh^{\mu\rho}\gh^\nu_\rho - \frac{\bR}{4}\, \gh_{\mu\nu}^2
    + \frac{\Lambda}{2}\, \gh_{\mu\nu}^2+ \mathcal{O}\left( \gh^3\right)\Bigg\rbrace
    \,,\\
    \mathcal{S}_\text{GB} = 
    \frac{\alpha\Mpl^2}{2}\int\!\!\dd^{d+4}x\sqrt{-\bg}\Bigg\lbrace
    & \nn 
    \bar{\mathcal{G}}
    -2\, \bigg[\bR^{\mu\alpha\beta\gamma}\bR^\nu_{\ \alpha\beta\gamma}
    -2\,\bR^{\mu\alpha\nu\beta}\bR_{\alpha\beta}
    -2\,\bR^{\mu\alpha}\bR^\nu_\alpha
    + \bR\,\bR^{\mu\nu}
    \bigg]\,\gh_{\mu\nu}\\
    &\bigg. \nn 
    + \bR^{\mu\rho\nu\sigma}\,\bn^\alpha\gh_{\rho\sigma}\bn_\alpha\gh_{\mu\nu}
    -2\,\bR^{\alpha\mu\rho\nu}\,\bn_\alpha\gh_{\rho\sigma}\bn^\sigma\gh_{\mu\nu}
    \\
    &\bigg. \nn 
    +6\,\bR^{\alpha\beta\mu\rho}\,\bn_\alpha\gh^\nu_\rho\bn_\beta\gh_{\mu\nu}
    -2\,\bR^{\mu\alpha\rho\beta}\,\bn_\alpha\gh^\nu_\rho\bn_\beta\gh_{\mu\nu}\\
    & \bigg. \nn 
    + \bR^{\alpha\beta}\,\bn_\alpha\gh_{\mu\nu}\bn_\beta\gh^{\mu\nu}
    +2\,\bR^{\mu\rho}\,\bn^\alpha\gh_\rho^\nu\bn_\alpha\gh_{\mu\nu}
    -\bR^{\mu\rho}\,\bn^\nu\gh_{\rho\sigma}\bn^\sigma\gh_{\mu\nu}\\
    & \bigg. \nn 
    -12\,\bR^{\mu\alpha}\,\bn_\alpha\gh_\rho^\nu\bn^\rho\gh_{\mu\nu}
    - \frac{\bR}{2}\,\bn_\alpha\gh_{\mu\nu}\bn^\alpha\gh^{\mu\nu}
    +3 \bR \,\bn^\mu\gh^\nu_\rho\bn^\rho\gh_{\mu\nu}\\
    & \bigg. \nn 
    + \bigg[ 
    \frac{3}{2}\,\bR^{\alpha\beta\mu\rho}\bR_{\alpha\beta}^{\quad\nu\sigma}
    -2\,\bR^{\alpha\beta\mu\rho}\bR_{\alpha\ \beta}^{\ \nu\ \sigma}
    -\bR^{\alpha\mu\beta\nu}\bR_{\alpha\ \beta}^{\ \rho\ \sigma}
    -4\,\bR^{\alpha\mu\beta\rho}\bR_{\alpha\ \beta}^{\ \nu\ \sigma}\\
    & \bigg. \nn  \qquad 
    +5\,\bR^{\alpha\mu\beta\rho}\bR_{\alpha\ \beta}^{\ \sigma\ \nu}
    +6\,\bR^{\alpha\mu\rho\nu}\bR_\alpha^\sigma
    -2\,\bR\,\bR^{\mu\rho\nu\sigma}
    +\bR^{\mu\nu}\bR^{\rho\sigma}
    -\bR^{\mu\rho}\bR^{\nu\sigma}
    \bigg]\,\gh_{\mu\nu}\gh_{\rho\sigma}\\
    & \bigg.\nn 
    + \bigg[
    4\,\bR^{\alpha\beta\gamma\mu}\bR_{\alpha\beta\gamma}^{\quad \ \nu} 
    -8\,\bR^{\alpha\mu\beta\nu}\bR_{\alpha\beta}
    -8\,\bR^{\alpha\mu}\bR_\alpha^\nu
    +4\,\bR\,\bR^{\mu\nu} 
    \bigg]\,\gh_\nu^\rho\gh_{\mu\rho}\\
    & \bigg.
    - \frac{1}{4}\,\bar{\mathcal{G}}\,\gh_{\mu\nu}\gh^{\mu\nu} 
    + \mathcal{O}\left( \gh^3\right) \Bigg\rbrace\,.
    \end{align}
\end{subequations}

\subsection{Projection onto the sub-manifolds with the original compactification scenario}\label{app_pert_scenario1}

Let us project the quadratic action onto each sub-manifold, using the compactification scenario~\eqref{eq_compactification_ansatz}.
The background Christoffel symbol is given by
\begin{subequations}
    \begin{align}
        &
        \bGa^0_{ij} = H_0\,a^2\, \hat{\delta}_{ij}\,,
        & &
        \bGa^i_{0j} = H_0\, \delta^i_j\,,\\
        &
        \bGa^i_{jk} = \frac{\hat{\delta}^{il}}{2}\big( \partial_j \hat{\delta}_{kl} + \partial_k \hat{\delta}_{jl} - \partial_l \hat{\delta}_{jk} \big)\,,
        & &
        \bGa^A_{BC} = \frac{\omega^{AD}}{2}\big( \partial_B \omega_{CD} + \partial_C \omega_{BD} - \partial_D \omega_{BC} \big)\,,
    \end{align}
\end{subequations}
all other components being null.
Hence, the non-vanishing components of the background Riemann and Ricci tensors are, up to the usual permutations,
\begin{subequations}
    \begin{align}
        &
        \bR_{0i0j} = - H_0^2\,a^2\,\hat{\delta}_{ij}\,,
        & &
        \bR_{ikjl} = H_0^2\,a^4\big( \hat{\delta}_{ij} \hat{\delta}_{kl} - \hat{\delta}_{il} \hat{\delta}_{jk} \big)\,,
        & &
        \bR_{ACBD} = \kappa \,b_0^2\,\big(\omega_{AB}\omega_{CD}-\omega_{AD}\omega_{BC}\big)\,,\\ 
        &
        \bR_{00} =-3\,H_0^2\,,
        & &
        \bR_{ij} = 3\,H_0^2\,a^2\,\hat{\delta}_{ij} \,,
        & &
        \bR_{AB} = (d-1)\,\kappa\, \omega_{AB} \,.
    \end{align}
\end{subequations}
The background Ricci and Gauss-Bonnet scalars hence reads
\begin{subequations}
    \begin{align}
    &
    \bR = 12 \, H_0^2 + d(d-1)\,\frac{\kappa}{b_0^2}\,,\\
    &
    \bar{\mathcal{G}} = 24\,H_0^4+24\,d(d-1)\,\frac{\kappa\,H_0^2}{b_0^2} + d(d-1)(d-2)(d-3)\,\frac{\kappa^2}{b_0^4}\,.
    \end{align}
\end{subequations}
As for the projections of the covariant derivatives, they are given by
\begin{subequations}
    \begin{align}
    &
    \bn_0\,\gh_{ij} = a^2\,\partial_t h_{ij}\,,
    & &
    \bn_0\,\gh_{AB} = b_0^2\,\partial_t H_{AB}\,,
    \\
    &
    \bn_k\,\gh_{ij} = a^2\,\hat{\partial}_k h_{ij}\,,
    & &
    \bn_i\,\gh_{AB} = b_0^2\,\partial_iH_{AB}\,,
    \\
    &
    \bn_A\,\gh_{ij} = a^2\,\partial_A h_{ij}\,,
    & &
    \bn_C\,\gh_{AB} = b_0^2\,\eth_CH_{AB}\,.
    \end{align}
\end{subequations}

Injecting those in the quadratic action, the linear term disappears as expected due to the TT properties of both $h_{ij}$ and $H_{AB}$. As for the quadratic term, after a last integration by part to tackle cross-derivative terms such as $\eth_C H_{AB}\,\eth_B H_{AC}$, it comes
\begin{subequations}
    \begin{align}
    \mathcal{S}^{(2)}_\text{EH} = \frac{\Mpl^2}{8}\int\!\!\dd t\, \dd^3x\,\dd^dx\,\sqrt\omega\,a^3b_0^d\,\Bigg\lbrace
    & \nn
    \big(\partial_t h_{ij}\big)^2- \frac{\big(\hp_k h_{ij}\big)^2}{a^2} -\frac{\big(\partial_A h_{ij}\big)^2}{b_0^2}\\
    &\nn
    + \big(\partial_t H_{AB}\big)^2- \frac{\big(\partial_i H_{AB}\big)^2}{a^2}- \frac{\big(\eth_C H_{AB}\big)^2}{b_0^2} \\
    &\nn
    + \bigg[2\Lambda -8H_0^2-d(d-1)\frac{\kappa}{b_0^2}\bigg]\,h_{ij}^2\\
    &
    + \bigg[2\Lambda -12H_0^2-\big(d^2-3d+4\big)\,\frac{\kappa}{b_0^2}\bigg]\,H_{AB}^2
    \Bigg\rbrace
    \,, \\
    \mathcal{S}^{(2)}_\text{GB} = \frac{\alpha\,\Mpl^2}{2}\int\!\!\dd t\, \dd^3x\,\dd^dx\,\sqrt\omega\,a^3b_0^d\,\Bigg\lbrace
    & \nn
    \frac{d(d-1)\kappa}{2\,b_0^2}\Bigg[\left(\partial_t h_{ij}\right)^2 - \left(\frac{\hp_k h_{ij}}{a}\right)^2\Bigg]\\
    &\nn
    - \Bigg[H_0^2+\frac{(d-1)(d-2)\kappa}{2\,b_0^2}\Bigg]\left(\frac{\partial_A h_{ij}}{b_0}\right)^2\\
    & \nn
    - \Bigg[\frac{d(d-1)(d-2)(d-3)\,\kappa^2}{4\,b_0^4}-8\,d(d-1)\,H_0^2\,\frac{\kappa}{b_0^2}-20\,H_0^4\Bigg]\,h_{ij}^2\\
    & \nn
    +\Bigg[\frac{(d-2)(d-3)\,\kappa}{2\,b_0^2}+3\,H_0^2\Bigg]\,\Bigg[\left(\partial_t H_{AB}\right)^2-\left(\frac{\partial_i H_{AB}}{a}\right)^2\Bigg]\\
    & \nn
    -\Bigg[\frac{(d-3)(d-4)\,\kappa}{2\,b_0^2}+6\,H_0^2\Bigg]\,\left(\frac{\eth_K H_{AB}}{b_0}\right)^2\\
    & \nn
    -\Bigg[\frac{(d-3)(d-4)(d^2-3d+6)\,\kappa^2}{4\,b_0^4}\\
    & \hspace{4cm}
    +6\big(d^2-3d+4\big)\,H_0^2\,\frac{\kappa}{b_0^2}+6\,H_0^4\Bigg]\,H_{AB}^2\Bigg\rbrace\,.
    \end{align}
\end{subequations}
Writing this action in terms of the dimensionless variables~\eqref{eq_adim_var} gives the result presented in Eqs.~\eqref{eq_d2S_tens} and~\eqref{eq_d2S_tens_detail}.

\subsection{Projection onto the sub-manifolds with the second compactification scenario}\label{app_pert_scenario2}

Let us now project the quadratic action onto each sub-manifold, using the second compactification scenario~\eqref{eq_compactification_ansatz_k0}.
For simplicity, we use the shorthand $L_0 \equiv H_0 \mathcal{X}_0$.
The background Christoffel symbol is given by
\begin{subequations}
    \begin{align}
        &
        \bGa^0_{ij} = H_0\,a^2\, \hat{\delta}_{ij}\,,
        & &
        \bGa^i_{0j} = H_0\, \delta^i_j\,,\\
        &
        \bGa^0_{AB} = L_0\,b^2\, \omega_{AB}\,,
        & &
        \bGa^A_{0B} = L_0\, \delta^A_B\,,\\
        &
        \bGa^i_{jk} = \frac{\hat{\delta}^{il}}{2}\big( \partial_j \hat{\delta}_{kl} + \partial_k \hat{\delta}_{jl} - \partial_l \hat{\delta}_{jk} \big)\,,
        & &
        \bGa^A_{BC} = \frac{\omega^{AD}}{2}\big( \partial_B \omega_{CD} + \partial_C \omega_{BD} - \partial_D \omega_{BC} \big)\,,
    \end{align}
\end{subequations}
all other components being null.
Hence, the non-vanishing components of the background Riemann and Ricci tensors are, up to the usual permutations,
\begin{subequations}
    \begin{align}
        &
        \bR_{0i0j} = - H_0^2\,a^2\,\hat{\delta}_{ij}\,,
        & &
        \bR_{0A0B} = - L_0^2\,b^2\,\omega_{AB}\,,\\
        & 
        \bR_{ikjl} = H_0^2\,a^4\big( \hat{\delta}_{ij} \hat{\delta}_{kl} - \hat{\delta}_{il} \hat{\delta}_{jk} \big)\,,
        & &
        \bR_{ACBD} = L_0^2 \,b^4\,\big(\omega_{AB}\omega_{CD}-\omega_{AD}\omega_{BC}\big)\,,\\ 
        &
        \bR_{iAjB} = H_0\,L_0\,a^2\,b^2\,\hat{\delta}_{ij} \omega_{AB}\,,
        & &
        \bR_{00} =-3\,H_0^2-d\,L_0^2\,,\\
        & 
        \bR_{ij} = (3\,H_0+d\,L_0)\,H_0\,a^2\,\hat{\delta}_{ij} \,,
        & &
        \bR_{AB} = (3\,H_0+d\,L_0)\,L_0\,b^2\, \omega_{AB}\,.
    \end{align}
\end{subequations}
The background Ricci and Gauss-Bonnet scalars hence reads
\begin{subequations}
    \begin{align}
    &
    \bR = 12 \, H_0^2 + 6d\,H_0,L_0 d(d+1)\,L_0^2\,,\\
    & \nn
    \bar{\mathcal{G}} = 24\,H_0^4 + 72\,d\,H_0^3\,L_0+24\,d(2d-1)\,H_0^2\,L_0^2\\
    & \qquad\qquad
    + 12\,d^2(d-1)H_0\,L_0^3 + (d+1)d(d-1)(d-2)\,L_0^4\,.
    \end{align}
\end{subequations}
As for the projections of the covariant derivatives, they are given by
\begin{subequations}
    \begin{align}
    &
    \bn_0\,\gh_{ij} = a^2\,\partial_t h_{ij}\,,
    & &
    \bn_0\,\gh_{AB} = b^2\,\partial_t H_{AB}\,,
    \\
    &
    \bn_k\,\gh_{ij} = a^2\,\hat{\partial}_k h_{ij}\,,
    & &
    \bn_i\,\gh_{AB} = b^2\,\partial_iH_{AB}\,,
    \\
    &
    \bn_A\,\gh_{ij} = a^2\,\partial_A h_{ij}\,,
    & &
    \bn_C\,\gh_{AB} = b^2\,\eth_CH_{AB}\,.
    \end{align}
\end{subequations}

Injecting those in the quadratic action, the linear term disappears as expected due to the TT properties of both $h_{ij}$ and $H_{AB}$. As for the quadratic term, both perturbations decouple, and it comes
\begin{subequations}
    \begin{align}
    \mathcal{S}^{(2)}_\text{EH} = \frac{\Mpl^2}{8}\int\!\!\dd t\, \dd^3x\,\dd^dx\,a^3b^d\,\Bigg\lbrace
    & \nn
    \big(\partial_t h_{ij}\big)^2- \frac{\big(\hp_k h_{ij}\big)^2}{a^2} -\frac{\big(\partial_A h_{ij}\big)^2}{b^2}\\
    &\nn
    + \big(\partial_t H_{AB}\big)^2- \frac{\big(\partial_i H_{AB}\big)^2}{a^2}- \frac{\big(\eth_C H_{AB}\big)^2}{b^2} \\
    &\nn
    + \bigg[2\Lambda -8H_0^2 -4d\,H_0\,L_0-d(d+1)\,L_0^2\bigg]\,h_{ij}^2\\
    &
    + \bigg[2\Lambda -12H_0^2-6(d-1)\,H_0\,L_0-\big(d^2-d+2\big)\,L_0\bigg]\,H_{AB}^2
    \Bigg\rbrace
    \,, \\
    \mathcal{S}^{(2)}_\text{GB} = \frac{\alpha\,\Mpl^2}{2}\int\!\!\dd t\, \dd^3x\,\dd^dx\,a^3b^d\,\Bigg\lbrace
    & \nn
    \Bigg[d\,H_0L_0+\frac{d(d-1)L_0^2}{2}\Bigg]\left(\partial_t h_{ij}\right)^2 
    - \frac{d(d-1)L_0^2}{2} \left(\frac{\hp_k h_{ij}}{a}\right)^2\\
    &\nn
    - \Bigg[H_0^2+(d-1)H_0L_0+\frac{d(d-1)L_0^2}{2}\Bigg]\left(\frac{\partial_A h_{ij}}{b}\right)^2\\
    & \nn
    +\Bigg[3H_0^2+3(d-2)H_0L_0+\frac{(d-2)(d-3)L_0^2}{2}\Bigg]\,\left(\partial_t H_{AB}\right)^2\\
    & \nn
    - \Bigg[3H_0^2+2(d-2)H_0L_0 + \frac{(d-1)(d-2)L_0^2}{2}\Bigg]\left(\frac{\partial_i H_{AB}}{a}\right)^2\\
    & \nn
    -\Bigg[6H_0^2+3(d-3)H_0L_0+\frac{(d-2)(d-3)L_0^2}{2}\Bigg]\,\left(\frac{\eth_K H_{AB}}{b}\right)^2\\
     & \nn
    + \Bigg[20\,H_0^4 + 38d\,H_0^3L_0+d(17d-3)H_0^2L_0^2\\
    & \nn \qquad\qquad
    +d^2(d-1)H_0\,L_0^3
    - \frac{(d+1)d(d-1)(d-2)\,L_0^4}{4}\Bigg]\,h_{ij}^2\\
    & \nn
    -\Bigg[6\,H_0^4+18(d-4)\,H_0^3L_0 + 6(2d^2-17d+9)\,H_0^2L_0^2\\
    & \qquad
    +3\big(d^3-13d^2+18d-2\big)\,H_0\,L_0^3 + \frac{(d-1)(d-2)(d^2-15d-4}{4}\Bigg]\,H_{AB}^2\Bigg\rbrace\,.
    \end{align}
\end{subequations}
Writing this action in terms of the dimensionless variables~\eqref{eq_adim_var} gives the result presented in Eqs.~\eqref{eq_d2S_tens_k0} and~\eqref{eq_d2S_tens_detail_k0}, with
\begin{subequations}\label{eq_constant_k0}
\begin{align}
    & \label{eq_noghost_phys_k0}
    \mathcal{K} = 
    \frac{2(1-\mathcal{X}_0)[1+(d-1)\mathcal{X}_0]}{2+(d-1)\mathcal{X}_0[4+(d-2)\mathcal{X}_0]}
    \,,\\
    & \label{eq_c2_phys_k0}
    c_\text{phys}^2 = 1 + \frac{d\,\mathcal{X}_0}{1+(d-1)\mathcal{X}_0}
    \,,\\
    & 
    c_\text{extr}^2 = 1 - \frac{1}{1+(d-1)\mathcal{X}_0}\,,\\
    &
    \mathcal{M}^2 =
    \frac{44+4(23d-2)\mathcal{X}_0+2(23d^2-9d+2)\mathcal{X}_0^2+6d^2(d-1)\mathcal{X}_0^3}{2+(d-1)\mathcal{X}_0[4+(d-2)\mathcal{X}_0]}
    \,,\\
    & \label{eq_noghost_extr_k0}
    \widetilde{\mathcal{K}} =
   -\frac{2(1-\mathcal{X}_0)[2+(d-2)\mathcal{X}_0]}{2+(d-1)\mathcal{X}_0[4+(d-2)\mathcal{X}_0]}
    \,,\\
    &
    \tilde{c}_\text{phys}^2 = 1 - \frac{(d-2)\mathcal{X}_0}{2+(d-2)\mathcal{X}_0}
    \,,\\
    &
    \tilde{c}_\text{extr}^2 = 1 + \frac{3}{2+(d-2)\mathcal{X}_0}
    \,,\\
    &
    \widetilde{\mathcal{M}}^2 = \frac{108\,\mathcal{X}_0+4(36d-17)\mathcal{X}_0^2+2(27d^2-35d-4)\mathcal{X}_0^3+2(d-1)(d-2)(3d+2)\mathcal{X}_0^4}{2+(d-1)\mathcal{X}_0[4+(d-2)\mathcal{X}_0]}
    \,,
\end{align}
\end{subequations}
where we used the background solution~\eqref{eq_bckg_sol_k0}.

\bibliography{ListRef_GB.bib}

\end{document}